\documentclass[aps,prb,preprint,floatfix,showpacs]{revtex4-1}
\usepackage{amsmath}
\usepackage{amsfonts}
\usepackage{amssymb}
\usepackage{graphicx}
\usepackage{hyperref}

\begin{document}

\title{Resonant Regimes in the Fock-Space Coherence of Multilevel
  Quantum dots}

\affiliation{Department of Physics and Atmospheric Science, Dalhousie
  University, Halifax, Nova Scotia, Canada, B3H 3J5}

\author{Eduardo Vaz}
\author{Jordan Kyriakidis}
\homepage{http://quantum.phys.dal.ca}

\begin{abstract}
  The coherence between quantum states with different particle numbers
  --- the Fock-space coherence --- qualitatively differs from the more
  common Hilbert-space coherence between states with equal particle
  numbers.  For a quantum dot with multiple channels available for
  transport, we find the conditions for decoupling the dynamics of the
  Fock-space coherence from both the Hilbert-space coherence as well
  as the population dynamics.  We further find specific energy and
  coupling regimes where a long-lived resonance in the Fock-space
  coherence of the system is realized, even where no resonances are
  found either in the populations or Hilbert-space coherence.
  Numerical calculations show this resonance remains robust in the
  presence of both boson-mediated relaxation and transport through the
  quantum dot.
\end{abstract}

\date{\today}


\maketitle

\section{Introduction}

According to quantum theory, a physical system can exist in a coherent
superposition of distinct values of a single physical observable.  One
such superposition is over the number of particles present in the
system.  Such Fock-space coherence (FSC) has been recently and
controllably demonstrated in bosonic
systems.~\cite{Bertet:2002p6290,Hofheinz:2008p5760,Bimbard:2010p6263}
The lifetime of these novel states may be considerably enhanced due to
the near quenching of single-particle relaxation channels.  This
mitigation of both decay and decoherence is a prime requisite, for
example, in the demonstration of scalable quantum information
processing architectures.  In Fermionic systems such as charge
carriers in a quantum dot, the influence of Fock states on the
dynamics and transport properties has been
investigated~\cite{datta_fock_2006,Harbola:2006p6149,Moldoveanu:2009p6148}
but not the coherence properties of the states themselves.  Here we
show that coherent fermionic Fock states in multilevel quantum dots
can exhibit decoherence times at least an order of magnitude
greater~\cite{FSC_orderOfMagnitude} than the more conventional
Hilbert-space coherence (HSC) between states with fixed particle
number.  This estimate remains valid even in the presence of a strong
perturbation such as resonant tunnelling transport through the system
as well as coupling to a bosonic bath.  Our fully non-Markovian
calculations reveal a memory-kernel whose elements factorise into
non-interacting blocks, one of which contains all of the
Fock-coherence.  Our results are independent of the number of
tunnelling pathways available through the system, although more than
one seems necessary, indicating that an interference effect may be
responsible for the robustness of these states.  Our results further
demonstrate how novel quantum states can exhibit remarkably useful
properties given careful---though not prohibitive---tuning of bias,
level spacing, and barrier anisotropy of the system.

\section{Model}

Our model is that of a single quantum dot weakly coupled to biased
source and drain semi-infinite leads, such that only a small number of
electronic channels are available for transport.  We also consider
particle-preserving relaxation through a coupling to a bosonic
reservoir (phonons, for example).  The total Hamiltonian is given by
\begin{equation} \label{eq:HTotal} H = H_{\text{leads}} + H_{\text{qd}} +
  H_{\text{boson}} + \mathcal{H}_{\mathcal{T}} +
  \mathcal{H}_{\mathcal{R}},
\end{equation}
where $\displaystyle H_{\text{leads}} = \sum_{s(d)} (\epsilon_{s(d)}
\pm \frac{1}{2} eV_b) d^\dag_{s(d)} d_{s(d)}$ describes the source and
drain leads as non-interacting fermion systems shifted by the bias
voltage $V_{\text{b}}$.  The creation and annihilation operators for
the source (drain) leads are $d^\dag_{s(d)}, d_{s(d)}$, respectively,
and $\epsilon_{s(d)}$ are the respective single-particle energies.
The quantum dot in the non-interacting regime is given by
$\displaystyle H_{\text{qd}} = \sum_i (\hbar\omega_i + eV_g) c^\dag_i
c_i$, where $c^\dag_i$ and $c_i$ are system creation and annihilation
operators, and the single-particle energies $\hbar \omega_i$ are all
shifted by the applied gate voltage, $V_g$.  For the sake of clarity
we neglected Coulomb interactions throughout the paper, as these are
not expected to qualitatively change the results beyond an
energy-level renormalization.~\cite{zero_coulomb,Moldoveanu:2010} The
coupling to the electronic reservoirs is described by a tunnelling
Hamiltonian, $\displaystyle \mathcal{H}_{\mathcal{T}} = \sum_{i, r =
  (s, d)} \left( T^r_i d^\dag_r c_i + \text{h.\,c.} \right)$, where
$T^{s(d)}_i$ is an energy-independent tunnelling coefficient for a
particle tunnelling from the single-particle state $|i\rangle$ in the
dot to the source (drain) reservoir~\cite{tunnel_index}.  We consider
the qualitative effects of a boson-mediated relaxation on the FSC, and
introduce a bosonic reservoir and its interaction with the quantum
dot, described respectively by $\displaystyle H_{\text{boson}}=
\sum_l\varepsilon_lb_l^\dag b_l$, and $\displaystyle
\mathcal{H}_{\mathcal{R}}=\sum_{i,j,l}\left(A_{ijl}c^\dag_ic_jb_l +
  \text{h.c.} \right)$, where $A_{ijl}$ is a generic coupling
coefficient containing the details of a specific electron-boson
interaction.

\section{Time evolution and the memory kernel}\label{kernel}
The non-Markovian time evolution of the quantum dot is considered in
the weak coupling approximation (Born approximation), where terms
up to second order in the interaction Hamiltonians
$\mathcal{H}_{\mathcal{T}}$ and $\mathcal{H}_{\mathcal{R}}$ are kept.

The generalized master equation for the reduced density
matrix~\cite{PhysRevB.81.085315} is given by
\begin{equation}
  \label{eq:rho_dot_born}
 \dot \rho_{ab}(t) = \sum_{c,d} \int_0^t \! dt'
  \rho_{cd}(t') \Upsilon_{abcd}(t-t') e^{i\gamma_{abcd} t'},
\end{equation}
where the total memory kernel $\Upsilon_{abcd}(t)$ describing the dynamics
of the system can be explicitly derived~\cite{unpub} directly from the
microscopic Hamiltonian Eq.~(\ref{eq:HTotal}).  Under a weak
coupling approximation $\Upsilon_{abcd}(t)$ can be written as the
sum of a transport-dependent ($\mathcal{T}$) and relaxation-dependent
($\mathcal{R}$) transition tensors,
\begin{equation}
  \label{eq:kernel}
  \Upsilon_{abcd}(t) = \mathcal{T}_{abcd}(t) + \mathcal{R}_{abcd}(t).
\end{equation}
The absence of bilinear terms in the interaction Hamiltonians
$\mathcal{H}_{\mathcal{T}}$ and $\mathcal{H}_{\mathcal{R}}$ is due to
the trace over the electronic reservoirs, where terms of the form
$\langle r| d_l | r\rangle=\langle r| d^\dag_l | r\rangle=0$.

The transition tensor $\mathcal{T}$ due to electronic transport is given
by,
 \begin{subequations}
   \label{eq:memory_kernel}
   \begin{align}
     \label{eq:R}
     \nonumber \mathcal{T}_{abcd}(t) = \sum_{\alpha,\beta,r} & \left\{
       K^r_{\alpha\beta} \left[ \Omega^{\alpha}_{\phi_B,\mu^r}(t)
         \Delta^{\alpha\beta}_{badc} -
         \Omega^{\alpha}_{\phi_T,\mu^r}(t)
         \Delta^{\alpha\beta}_{cdab} \right] \right. \\
     + &\left. K^r_{\beta\alpha} \left[ \Omega^{\alpha \
           *}_{\phi_B,\mu^r}(t) \Delta^{\alpha\beta}_{abcd} -
         \Omega^{\alpha \ *}_{\phi_T,\mu^r}(t)
         \Delta^{\alpha\beta}_{dcba} \right] \right\},
   \end{align}
   with,
   \begin{gather}
     \label{eq:Omega}
     \Omega^{\alpha}_{x,y}(t) = \left( \text{e}^{i \omega_{x\alpha} t}
       -
       \text{e}^{i \omega_{y\alpha} t}\right) / t,\\
     \label{eq:K}
     K^r_{\alpha \beta} \equiv i N_r \left( T^r_{\alpha}\right)^*
     T_{\beta}^r / \hbar,\\
    \label{eq:Delta}
    \Delta^{\alpha \beta}_{abcd} \equiv \langle a | c^{\dag}_{\alpha}|
    c \rangle \langle d | c_{\beta} | b \rangle - \langle a | c
    \rangle \langle b | c_{\alpha} c^{\dag}_{\beta} | d \rangle,
   \end{gather}
 \end{subequations}
 and where the Latin indexes denote many-body system states, Greek
 indexes denote single-particle states, and $r = s,\ d$ denotes the
 source and drain leads.  Both leads are assumed to be metallic with
 occupied-state energies lying between a lower bound $\phi_{B}$ and
 the lead's chemical potential $\mu^{s(d)}$, and unoccupied-state energies
 lying between the lead's chemical potential and an upper bound
 $\phi_{T}$.  The range between the bounds $\phi_{B}$ and $\phi_{T}$
 are taken to be large relative to the characteristic energies of the
 system.  Finally, $N_r$ denotes the density of states in lead $r$,
 and the frequencies $\omega_{ij} \equiv \omega_i - \omega_j$.

 Similarly, the transition tensor due to boson-mediated relaxation is given by
 \begin{subequations}
   \label{eq:boson_kernel}
   \begin{align}
     \label{eq:Q}
     \nonumber \mathcal{R}_{abcd}(t) = G \sum_{ijpqk} &\left[ M_{ijpqk} \left(
         \Theta^{abcd}_{ijpq} \text{e}^{i \omega_{ijk} t} +
         \Theta^{abcd}_{pqij} \text{e}^{-i \omega_{ijk} t} \right)  \right. \\
     + &\left. M^*_{ijpqk} \left( \Theta^{dcba}_{ijpq} \text{e}^{i
           \omega_{ijk} t} + \Theta^{dcba}_{pqij} \text{e}^{-i
           \omega_{ijk} t} \right) \right]
   \end{align}
   where,
   \begin{gather}
     \label{eq:boson_kernel_defs}
     \Theta_{ijpq}^{abcd} = \langle a | c_i^\dag c_j | c \rangle
     \langle d | c_q c_p^\dag | b \rangle - \langle b | d \rangle
     \langle a |
     c_q^\dag c_p  c_j c_i^\dag | c \rangle,  \\
     M_{ijpqk} = A_{ijk} A^*_{pqk},
   \end{gather}
 \end{subequations}
 and $G$ is the density of states of the boson reservoir.

\section{Two transport channels}\label{2CH}

The transport and relaxation transition tensors presented in
Eqs.~(\ref{eq:memory_kernel}, \ref{eq:boson_kernel}), are for a
general number of available transport channels.  For $k$ transport
channels, a minimum of $2^k$ dynamical states are required, yielding
$2^{2k}$ density matrix elements and $2^{4k}$ transition tensor
components determining the dynamics of the system; despite
symmetry-induced reductions in the number of independent components,
the computational cost nevertheless increases exponentially with the
number of channels.  In what follows, we focus on the simplest
non-trivial case of two available transport channels.

\begin{figure}
\begin{center}
  \includegraphics[width=24pc]{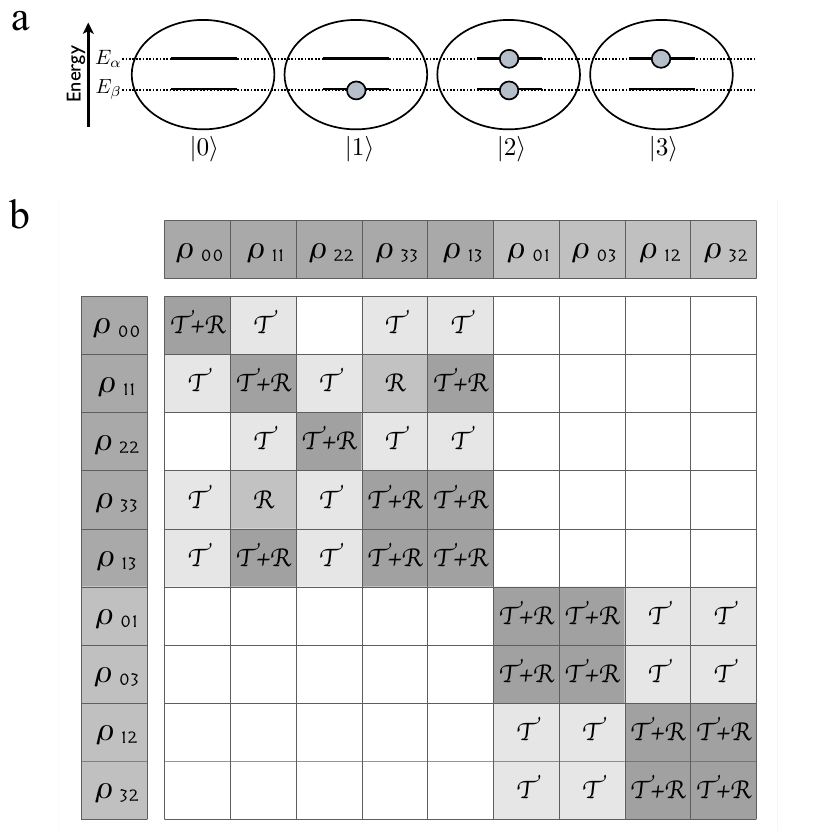}
  \caption{\label{fig:2CH} (a) Basis states for the two-channel case,
    defined as $|0\rangle = |N-1 \rangle_{\text{gs}}$, $|1\rangle =
    |N\rangle_{\text{gs}}$, $|2\rangle = |N + 1\rangle_{\text{gs}}$,
    and $|3\rangle = |N\rangle_{\text{es}}$.  (b) Representation of
    the non-zero memory kernel elements for the case of two transport
    channels.  In the figure, the matrix elements represent the
    contributions to the dynamics between the reduced density matrix
    elements shown in the outermost row and column, due to transport
    ($\mathcal{T}$), bosonic relaxation ($\mathcal{R}$), or both.  Not
    all elements of $\Upsilon_{abcd}$ are depicted; missing elements
    are the complex conjugates of those present.}
\end{center}
\end{figure}

The two-channel model is schematically depicted in
Fig.~\ref{fig:2CH}a.  States $|0\rangle$, $|1\rangle$, and $|2\rangle$
respectively denote ground states with $N-1$, $N$, and $N+1$ confined
particles and state $|3\rangle$ denotes an $N$-particle excited state.
This is an experimentally accessible
regime.~\cite{Kyriakidis2002Voltage-tunable} The non-vanishing
components of the memory kernel are shown in Fig.~\ref{fig:2CH}b.  The
cell in the row labelled $\rho_{ab}$ and column $\rho_{cd}$ indicates
whether the element $\Upsilon_{abcd}$ is zero (empty cell) or depends
on a combination of transport ($\mathcal{T}$) or boson-mediated
relaxation ($\mathcal{R}$).  For example, only transport contributes
to $\Upsilon_{3300}$, only bosonic relaxation contributes to
$\Upsilon_{1133}$, both transport and bosonic relaxation contribute to
$\Upsilon_{1113}$, and $\Upsilon_{0022}$ vanishes.  Elements not
depicted in the figure may be inferred through the relation
$\Upsilon_{abcd} = \Upsilon^*_{badc}$ for $a \neq b$ and $c \neq
d$.~\cite{rho02} As expected, the only transitions where relaxation
alone ($\mathcal{R}$) carries the dynamics are between states with the
same number of particles ($|1\rangle$ and $|3\rangle$ in this case).

In this matrix representation of the tensor $\Upsilon_{abcd}$, two
decoupled sub-matrices are apparent.  The upper sub-matrix represents
dynamics among population probabilities and coherence in states with
identical particle number (HSC).  The lower block describes coherence
between states differing by one particle.  Thus, the lower block
describes the evolution of the FSC and is decoupled from the HSC.

The decoupling of the FSC dynamics is not a result of the number of
channels involved.  For example, increasing the bias to allow more
transport channels will increase the number of relevant elements of
$\Upsilon_{abcd}$.  The structure that emerges is similar to that
depicted in Fig.~\ref{fig:2CH}b, but the blocks themselves increase in
both size and number; the FSC blocks remain decoupled from the HSC.
Fundamentally, the structure is due to the fact that fermions cannot
be individually created or destroyed through a scattering event (as
opposed to bosonic systems).  Since a bosonic bath does not alter the
structure of the transition tensor, and since the time scale set by
resonant tunnelling is generally much faster than that set by the
bosonic coupling, we shall restrict ourselves to the transport problem
and neglect the bosons in the remainder.

A solution to the evolution equations, Eq.~(\ref{eq:rho_dot_born}), is
obtained~\cite{PhysRevB.81.085315} by moving to Laplace space and
using the convolution theorem to transform the integrodifferential
equations to a coupled set of algebraic equations.  We numerically
uncouple these equations and then transform back to the time domain by
means of a Bromwich integral~\cite{integration}.

The present transport model is described by seven parameters: Fermi
energy $E_{\text{F}}$, bias voltage $V_{b}$, gate voltage $V_g$,
energy-level spacing $\Delta E = E_{\alpha} - E_{\beta}$, orbital
anisotropy
$\varepsilon=1-T_{\beta}^{\text{s}}/T_{\alpha}^{\text{s}}=1-T_{\beta}^{\text{d}}/T_{\alpha}^{\text{d}}$,~\cite{Tequal}
barrier asymmetry $\lambda =
T_{\alpha}^{\text{d}}/T_{\alpha}^{\text{s}}=
T_{\beta}^{\text{d}}/T_{\beta}^{\text{s}}$, and overall
tunnel-coupling strength set by $T_{\alpha}^{\text{s}}$.  The orbital
anisotropy $\varepsilon$ describes differences in the tunnel coupling
strength to different orbitals in the system.  We define \emph{edge}
and \emph{core} orbitals by the relative strength of their coupling to
the reservoirs owing to the detailed shape of the wave function and
the relative penetration into the tunnelling
barriers.~\cite{ciorg02:collap.spin.singl} In relation to the
2-channel case illustrated in Fig.~\ref{fig:2CH}a, we denote orbitals
$|\alpha\rangle$ and $|\beta\rangle$ as the edge and core orbitals
respectively.

\begin{table*}[t]
  \begin{center}
\begin{tabular}{ c || c  c  c  c  c  c  c }
 Parameter & $E_F$ & $eV_B$ & $eV_g$ & $\Delta E$ & $\varepsilon$ &
  $\lambda$ & $T^{\text{s}}_{\alpha}$ \\
\hline
Value & $\ 30 \ \text{meV} \ $ & $\ 6 \ \text{meV} \ $ & $\ 0 \ \text{meV} \ $ & $\ 2 \ \text{meV}
\ $ & $\ \ \ 1 \ \ \ $
& $\ \ \ 1 \ \ \ $ & $\ 0.5 \ \text{meV} \ $ \\
\end{tabular}
\end{center}
 \caption{Standard set of parameter values used 
throughout the present work.  Unless otherwise 
noted, these values were used for all results 
presented. }
\label{tab:1}
\end{table*}

In what follows, we plot our results as a function of one or two parameters while 
fixing the others to a set of ``standard values."  The standard values of the 
parameters are given in Table~\ref{tab:1}, which exhaustively describes our energetic 
configuration.  For example, the two orbital levels have an energy of
$E_{\alpha,\beta}=E_{\text{F}}\pm \frac{\Delta E}{2} + eV_g$.  

\section{FSC resonance}\label{FSCR}

\begin{figure}
  \begin{center}
    \includegraphics[width=24pc]{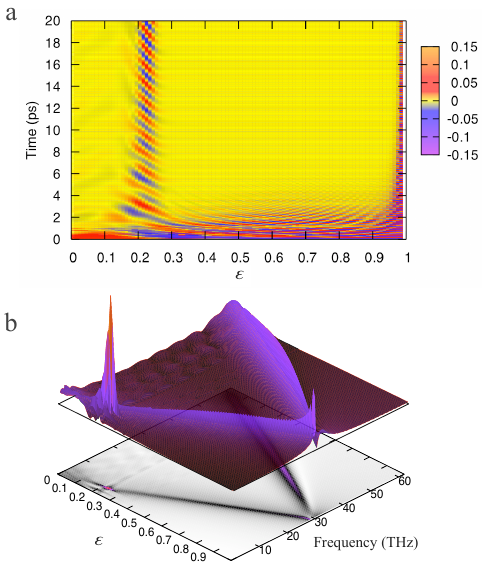}
    \caption{\label{fig:OA_FFT} (a) Time evolution of the FSC element
      $\rho_{01}(t)$ as a function of the orbital anisotropy
      $\varepsilon$.  A resonance in the FSC is observed at an
      anisotropy of $\varepsilon_{\text{res}} \approx 0.22$ that
      maintains a persistent coherence well beyond the expected
      decoherence time.  (b) Fourier transform of the results in (a).
      A linear dependence between frequency and $\varepsilon$ is
      evident.  A spike in the low frequency amplitude occurs at the
      resonant anisotropy of $\varepsilon=0.22$.  From this peak, we
      obtain the resonant frequency as given by
      Eq.~(\ref{eq:res_frequency}).}
  \end{center}
\end{figure}

We focus on the density-matrix element $\rho_{01}(t)$ which describes
the FSC between the $N - 1$ and the $N$-particle ground states.  (See
Fig.~\ref{fig:2CH}a.)  The initial conditions for the density-matrix
have an effect only on the amplitude of oscillations of the FSC, and
not on the presence or location of a resonance.  Thus, for the results
presented here we choose a homogeneous initial FSC distribution as
$\rho_{01}(0)=\rho_{03}(0)=\rho_{12}(0)=\rho_{32}(0)= 0.1$.

The time evolution of the FSC element $\rho_{01}(t)$ is shown in
Fig.~\ref{fig:OA_FFT}a as a function of orbital anisotropy
$\varepsilon$ (horizontal axis), where $T_{\beta}=\chi T_{\alpha}$
with $0 \le \chi \le 1$. For all but a narrow range around
$\varepsilon_{\text{res}} \approx 0.22$, the coherence drops off
rapidly, within approximately 3~ps, corresponding to only a few
tunnelling events ($\hbar / T_\alpha \approx 0.7$~ps).  This mirrors
the decay found in the HSC,~\cite{PhysRevB.81.085315} described, for
example, by the matrix element $\rho_{13}(t)$.  Remarkably, at an
orbital anisotropy of $\varepsilon_{\text{res}} \approx 0.22$, we
observe a resonance in the FSC.  The coherent oscillations persist for
a far greater length of time.  This resonance is unique to the FSC
elements of $\rho(t)$; they do not occur in either the HSC or the
population probabilities $\rho_{nn}(t)$.  To our knowledge, this is
the first demonstration of a resonance in the Fock-Space coherence of
an open quantum system of fermions.

Figure~\ref{fig:OA_FFT}b shows the Fourier spectrum of the time-domain
results shown in Fig.~\ref{fig:OA_FFT}a.  At a given $\varepsilon$,
two frequencies dominate the spectrum.  One is a high-frequency
component which decreases linearly with $\varepsilon$.  This is the
frequency related to the fast decay of the HSC, the population
probabilities, and the FSC at very short times.  There is also a
low-frequency component evident in Fig.~\ref{fig:OA_FFT}b,
corresponding to slow envelope oscillations.  The resonance, clearly
seen as the spike in Fig.~\ref{fig:OA_FFT}b, occurs on this
low-frequency line.

\begin{figure}
  \begin{center}
    \includegraphics[width=24pc]{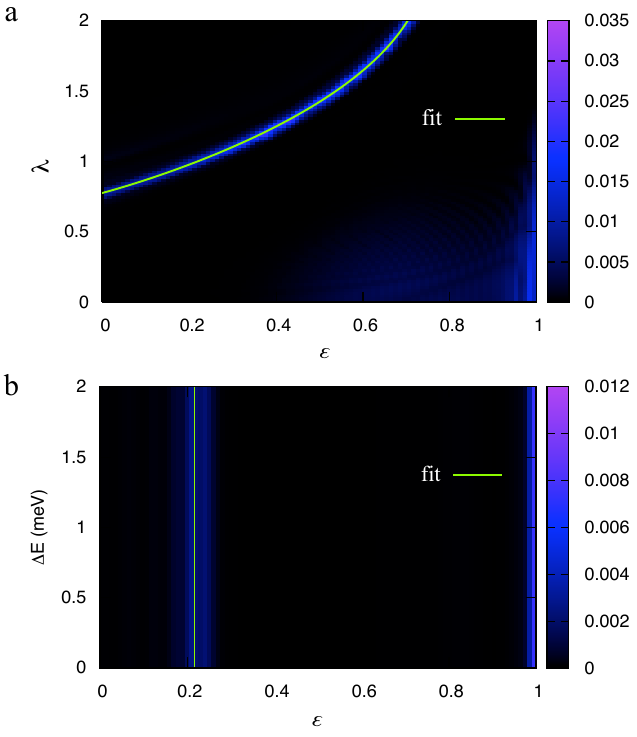}
    \caption{\label{fig:OA_coup} (a) FSC element $\rho_{01}$ in the
      long-time limit as a function of orbital anisotropy
      $\varepsilon$, and barrier asymmetry $\lambda$.  The ridge
      denotes the resonance.  (b) FSC element $\rho_{01}$ at the
      long-time limit as a function of orbital anisotropy and energy
      level spacing.  The fits displayed in both plots where produced
      according to Eq.~(\ref{eq:epsilon_resonant}).}
  \end{center}
\end{figure}

The resonant orbital anisotropy $\varepsilon_{\text{res}}$ depends on
the opacity of the source and drain barriers as well as on the overall
strength of the tunnel couplings.  Figure~\ref{fig:OA_coup}a shows
$\rho_{01}$ in the long-time limit (defined as the time beyond which
the transients have decayed---approximately 10~ps in the present case)
as a function of both orbital anisotropy $\varepsilon$ and barrier
asymmetry $\lambda$.  The evident peak (resonance) in the amplitude of
$\rho_{01}$ at this long-time limit gives the position of the resonant
orbital anisotropy as a function of the barrier asymmetry.  We observe
a strong relationship between $\lambda$ and $\varepsilon_{\text{res}}$
whose form is obtained from an empirical analysis (discussed later in
this paper), and shown superimposed on the calculated data on
Fig~\ref{fig:OA_coup}a.  A similar dependence is also obtained between
the resonant orbital anisotropy and the overall tunnelling strength.
In contrast, the dependence of $\varepsilon_{\text{res}}$ on the
energetics is rather weak.  Figure~\ref{fig:OA_coup}b, for example,
shows the FSC resonance, again in the long-time limit, as a function
of orbital anisotropy $\varepsilon$ and level spacing $\Delta E \equiv
E_{\alpha} - E_{\beta}$, and exhibits a linear and much weaker
dependence of $\varepsilon_{\text{res}}$ on $\Delta E$.  (Similar
results are seen in the dependence of $\varepsilon_{\text{res}}$ on
the bias voltage $V_{\text{b}}$, and gate voltage $V_{\text{g}}$.)

\begin{figure}
  \begin{center}
    \includegraphics[width=24pc]{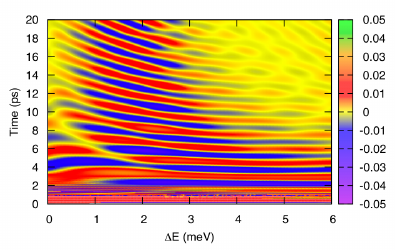}
    \caption{\label{fig:deltaE_res} Time evolution of the FSC element
      $\rho_{01}(t)$ as a function of the level spacing within the
      bias window, and for a resonant orbital anisotropy $\varepsilon=0.22$.}
  \end{center}
\end{figure}

As might be expected, the energetics of the system do play a role in
the frequency of the coherent oscillations.
Figure~\ref{fig:deltaE_res} plots the time evolution of the FSC
element $\rho_{01}(t)$ as a function of the orbital level spacing
$\Delta E$ for a resonant anisotropy $\varepsilon_{\text{res}} =
0.22$.  The patterns seen in Figure~\ref{fig:deltaE_res} are
suggestive of a coherent interference effect between tunnelling
pathways as the source of the resonance.  The strongest constructive
interference for this set of parameters occurs for $\Delta E \approx
1.5$~meV.  

We have empirically deduced that the dominant frequencies $\nu$
evident in Fig.~\ref{fig:OA_FFT}b depend on the energy and coupling
system parameters in the following way:
\begin{equation}
  \label{eq:frequency}
  \nu_{\pm}(\varepsilon) = \left| E_{\text{F}} + e V_g - \frac{\Delta
      E}{2} \pm 2 \text{ln}(\phi_T) (T_{\alpha}^s)^2(1+\lambda^2) \left(1 - \varepsilon\right) \right|.
\end{equation}
We have further empirically deduced that the resonant frequency
$\nu_{\text{res}}$, also evident in Fig.~\ref{fig:OA_FFT}b, depends on
the energetics of the system in the following way:
\begin{equation}
  \label{eq:res_frequency}
  \nu_{\text{res}} = \frac{1}{2} (e V_{\text{b}} + \Delta E) + e V_g.
\end{equation}
At resonance, Eqs.~(\ref{eq:frequency}) and (\ref{eq:res_frequency})
describe the same frequency.  Equating them yields the resonance
anisotropy $\varepsilon_{\text{res}}$ as a function of the Fermi
energy, level-spacing, bias voltage, gate voltage, coupling strength,
and barrier anisotropy:
\begin{equation}
  \label{eq:epsilon_resonant}
  \varepsilon_{\text{res}} = 1 - \left( \frac{ E_F -
    \Delta E - \frac{1}{2}eV_{\text{b}}}{2 \text{ln}(\phi_T) (T_{\alpha}^s)^2(1+\lambda^2) }\right).
\end{equation}

The fits to the numerical results presented in
Figs.~\ref{fig:OA_coup}a and \ref{fig:OA_coup}b were produced
according to Eq.~(\ref{eq:epsilon_resonant}).  Note that there are no
free fitting parameters in this result.  These result may be used as
an experimental guide when searching for this FSC resonance, as novel
methods to extract electronic THz oscillations (equivalent to
picosecond-range periods) are becoming accessible either by indirect
electrooptical methods~\cite{Auston_Nuss_1988, RevModPhys.83.543} or
by means of high-Tc superconductor
devices~\cite{Cavalleri_Nature_2011}.

\section{Source of FSC Resonance}

In order to address the source of the FSC resonance in our model, we
consider the formation of a bound state between the quantum dot and
the reservoirs.~\cite{Miyamoto_2005,Bulgakov_2007,Tong_2010}  Such a
bound state may arise when the presence of a coupling term between the
system and environment reduces the total energy.

Defining the bound state of the QD and the reservoirs as
$|\psi\rangle=|QD\rangle|R\rangle$, where $|QD\rangle=\sum_n
P_n|n\rangle$ denotes the state of the QD, and
$|R\rangle=\sum_r \int_{\phi^r_B}^{\phi^r_T} f(\epsilon_k)|\epsilon_k\rangle d\epsilon_k$ denotes the
continuum of reservoir states for $r=\{s,d\}$, we solve
the eigenvalue equation $H|\psi\rangle=E|\psi\rangle$ for the total
energy of the $E$, with the total Hamiltonian $H$ given by
Eq.~\ref{eq:HTotal}.

For a $2$-channel case, the solution to the eigenvalue equation has the
form,~\cite{Miyamoto_2005},
\begin{equation}
  \label{eq:eigeneq}
  \sum_{n'=0}^{3}[E_n\delta_{n,n'}-\sum_{r=\{s,d\}}\int_{\phi^r_B}^{\phi^r_T}
  \frac{T^r_{n'}(T^r_n)^*}{\epsilon_k-E}d\epsilon_k]P_{n'}=E P_n.
\end{equation}

We note that the integration limits over the reservoirs become
truncated by the single creation and annihilation operators of the
tunneling Hamiltonian to $\int_{\phi^r_B}^{E^r_{Fermi}}$ for removing
a particle from, and $\int^{\phi^r_T}_{E^r_{Fermi}}$ for adding a
particle to, reservoir $r$.

Therefore, the solution is found to be of the form,
\begin{equation}
  \label{eq:transcendental}
  Y(E)=E_0-I_{\alpha,\alpha,E_3}-I_{\beta,\beta,E_1}-\frac{[I_{\alpha,\beta,E_1}+
I_{\beta,\alpha,E_3}][I_{\alpha,\beta,E_3}+I_{\beta,\alpha,E_1}]}{E_2-E-[I_{\alpha,\alpha,E_1}+I_{\beta,\beta,E_3}]}=E,
\end{equation}
with,
\begin{equation}
  \label{eq:integrals}
  I_{a,b,E_n}=E_n - E -
  \sum_{r=\{s,d\}}\left[\int^{\phi^r_T}_{E^r_{Fermi}}
    \frac{T^r_a(T^r_b)^*}{\epsilon_k-E}d\epsilon_k +
    \int_{\phi^r_B}^{E^r_{Fermi}} 
    \frac{T^r_a(T^r_b)^*}{\epsilon_k-E}d\epsilon_k]\right].
\end{equation}

The existence of a bound state requires that $Y(E)$ has at minimum one
real solution for $E<0$, which is equivalent to $Y(0)<0$ being
satisfied.  For the case of the transcendental equation,
Eq.~\ref{eq:transcendental}, we find that the left hand side is
monotonically decreasing with $E$, and the right hand side linearly
increases with $E$, therefore there can be at most one solution where
l.h.s. = r.h.s., and thus only one bound state.  In the case of
$Y(0)<0$, we also see that under the approximation of
energy-independent coupling coefficients, the integrals over
$\epsilon_k$ appearing in Eq.~\ref{eq:integrals} lead to terms of the
form $ln(\epsilon_k)|_{\epsilon_k=\{\phi^r_{(B,T)}, E^r_{Fermi}\}}$.
Using this, along with recalling the coupling anisotropies
$\lambda=T^d_n/T^s_n$ and $\varepsilon= 1-T^r_\alpha/T^r_\beta$ in
order to define all coupling coefficients in terms of $T^s_\alpha=T$:
$T^s_\beta=(1-\varepsilon)T; T^D_\alpha=\lambda T;
T^D_\beta=(1-\varepsilon)\lambda T$, we find the following form for
$\varepsilon$,
\begin{equation}
  \label{eq:varepsilon2}
  \varepsilon \propto 1-
  \frac{F(E_0,E_1,E_2,E_3,E^s_{Fermi}, E^d_{Fermi})}{ln(\phi^s_T)T^2(1+\lambda^2)},
\end{equation}
where $F(E_0,E_1,E_2,E_3,E^s_{Fermi}, E^d_{Fermi})$ is an algebraic
function of all the energy parameters.  The important feature to point
out in Eq.~\ref{eq:varepsilon2} is the overall similarity with
Eq.~\ref{eq:epsilon_resonant}.  The fact that a similar relationship
between the coupling and anisotropy parameters of the system has been
found from two very different approaches is remarkable: the creation
of the relationship out of empirical analysis of the GME for the open
system, and the analytical derivation from a bound state analysis of
the eigenvalue equation.

The formation of the bound state can be linked to an interference
effect between transmission resonances due to the presence of several
transport channels in the system --- a feature eluded to in
Fig.~\ref{fig:deltaE_res}.  Resonances of this type have been observed
in diverse systems such as in quantum billiards,~\cite{Sadreev_2006}
laser induced continuum structures in atoms,~\cite{Magunov_1999} and
quantum dots.~\cite{Rotter_2005} In these works, it has been revealed
that the formation of bound states in the continuum is directly
related to the shape of the system and anisotropy of the coupling
parameters.  This point is specifically reflected in our findings as
we only observe the FSC resonance under coupling anisotropies.  These
coupling anisotropies may play the role of a path difference between
the channels, where specific path differences may quench transport
through the system in an effect akin to negative differential
conductance.~\cite{vaz_kyriakidis_2008} In such a case the FSC is
enhanced --- forming a resonance --- since the dominant decoherence
pathways are also effectively quenched.  This may also help explain
the presence of the single dominant resonance in the range of the
orbital anisotropy in our model, since only a specific energy and
coupling configuration will lead to transport quenching.

\section{Conclusions}

In conclusion, we have shown that the FSC of a fermionic system can be
decoupled from the evolution of the population probabilities and the
HSC of the system, even under the presence of boson-mediated
relaxation and single-particle exchange with a bath.  We demonstrated
a resonant regime in this FSC where the decoherence times extend far
beyond that of the HSC, and we have been able to relate these results
to an analytically derived relationship in the coupling parameters of
the system arising from the formation of a bound state between the
system and reservoirs.  We anticipate the primary difficulty in the
experimental confirmation of our theory to be the establishment of the
FSC in the first instance.  What we have shown here is that such a FSC
can exhibit a remarkable, robust, and long-lived resonance even in the
presence of ordinarily very destructive perturbations such as
particles tunnelling into and out of the system.  A plausible
configuration to obtain this FSC resonance effect is one where the
particle reservoirs are macroscopic systems in thermodynamic
equilibrium, as envisioned in the present work, and whose description
is properly and most conveniently characterised by a particle density
(not particle number) in a grand canonical ensemble.  Subsequent
experimental and further theoretical efforts extending this work may
shed light on the fundamental problem of decoherence, and on quantum
information processing in a semiconductor environment.

\begin{acknowledgments}
  This work is supported by NSERC of Canada, and by Lockheed Martin Corporation.
\end{acknowledgments}


\begin{thebibliography}{29}
\expandafter\ifx\csname natexlab\endcsname\relax\def\natexlab#1{#1}\fi
\expandafter\ifx\csname bibnamefont\endcsname\relax
  \def\bibnamefont#1{#1}\fi
\expandafter\ifx\csname bibfnamefont\endcsname\relax
  \def\bibfnamefont#1{#1}\fi
\expandafter\ifx\csname citenamefont\endcsname\relax
  \def\citenamefont#1{#1}\fi
\expandafter\ifx\csname url\endcsname\relax
  \def\url#1{\texttt{#1}}\fi
\expandafter\ifx\csname urlprefix\endcsname\relax\def\urlprefix{URL }\fi
\providecommand{\bibinfo}[2]{#2}
\providecommand{\eprint}[2][]{\url{#2}}

\bibitem[{\citenamefont{Bertet et~al.}(2002)\citenamefont{Bertet, Osnaghi,
  Milman, Auffeves, Maioli, Brune, Raimond, and Haroche}}]{Bertet:2002p6290}
\bibinfo{author}{\bibfnamefont{P.}~\bibnamefont{Bertet}},
  \bibinfo{author}{\bibfnamefont{S.}~\bibnamefont{Osnaghi}},
  \bibinfo{author}{\bibfnamefont{P.}~\bibnamefont{Milman}},
  \bibinfo{author}{\bibfnamefont{A.}~\bibnamefont{Auffeves}},
  \bibinfo{author}{\bibfnamefont{P.}~\bibnamefont{Maioli}},
  \bibinfo{author}{\bibfnamefont{M.}~\bibnamefont{Brune}},
  \bibinfo{author}{\bibfnamefont{J.~M.}~\bibnamefont{Raimond}}, \bibnamefont{and}
  \bibinfo{author}{\bibfnamefont{S.}~\bibnamefont{Haroche}},
  \bibinfo{journal}{Phys. Rev. Lett.} \textbf{\bibinfo{volume}{88}},
  \bibinfo{pages}{143601} (\bibinfo{year}{2002}).

\bibitem[{\citenamefont{Hofheinz et~al.}(2008)\citenamefont{Hofheinz, Weig,
  Ansmann, Bialczak, Lucero, Neeley, O'connell, Wang, Martinis, and
  Cleland}}]{Hofheinz:2008p5760}
\bibinfo{author}{\bibfnamefont{M.}~\bibnamefont{Hofheinz}},
  \bibinfo{author}{\bibfnamefont{E.~M.} \bibnamefont{Weig}},
  \bibinfo{author}{\bibfnamefont{M.}~\bibnamefont{Ansmann}},
  \bibinfo{author}{\bibfnamefont{R.~C.} \bibnamefont{Bialczak}},
  \bibinfo{author}{\bibfnamefont{E.}~\bibnamefont{Lucero}},
  \bibinfo{author}{\bibfnamefont{M.}~\bibnamefont{Neeley}},
  \bibinfo{author}{\bibfnamefont{A.~D.} \bibnamefont{O'connell}},
  \bibinfo{author}{\bibfnamefont{H.}~\bibnamefont{Wang}},
  \bibinfo{author}{\bibfnamefont{J.~M.} \bibnamefont{Martinis}},
  \bibnamefont{and} \bibinfo{author}{\bibfnamefont{A.~N.}
  \bibnamefont{Cleland}}, \bibinfo{journal}{Nature}
  \textbf{\bibinfo{volume}{454}}, \bibinfo{pages}{310} (\bibinfo{year}{2008}).

\bibitem[{\citenamefont{Bimbard et~al.}(2010)\citenamefont{Bimbard, Jain,
  MacRae, and Lvovsky}}]{Bimbard:2010p6263}
\bibinfo{author}{\bibfnamefont{E.}~\bibnamefont{Bimbard}},
  \bibinfo{author}{\bibfnamefont{N.}~\bibnamefont{Jain}},
  \bibinfo{author}{\bibfnamefont{A.}~\bibnamefont{MacRae}}, \bibnamefont{and}
  \bibinfo{author}{\bibfnamefont{A.}~\bibnamefont{Lvovsky}},
  \bibinfo{journal}{Nature Photon.} \textbf{\bibinfo{volume}{4}}
  (\bibinfo{year}{2010}).

\bibitem[{\citenamefont{Datta}(2006)}]{datta_fock_2006}
\bibinfo{author}{\bibfnamefont{S.}~\bibnamefont{Datta}},
  \emph{\bibinfo{title}{Fock space formulation for nanoscale transport}}
  (\bibinfo{year}{2006}), \eprint{arXiv:cond-mat/0603034}.

\bibitem[{\citenamefont{Harbola et~al.}(2006)\citenamefont{Harbola, Esposito,
  and Mukamel}}]{Harbola:2006p6149}
\bibinfo{author}{\bibfnamefont{U.}~\bibnamefont{Harbola}},
  \bibinfo{author}{\bibfnamefont{M.}~\bibnamefont{Esposito}}, \bibnamefont{and}
  \bibinfo{author}{\bibfnamefont{S.}~\bibnamefont{Mukamel}},
  \bibinfo{journal}{Phys.\ Rev.\ B} \textbf{\bibinfo{volume}{74}},
  \bibinfo{pages}{235309} (\bibinfo{year}{2006}).

\bibitem[{\citenamefont{Moldoveanu et~al.}(2009)\citenamefont{Moldoveanu,
  Manolescu, and Gudmundsson}}]{Moldoveanu:2009p6148}
\bibinfo{author}{\bibfnamefont{V.}~\bibnamefont{Moldoveanu}},
  \bibinfo{author}{\bibfnamefont{A.}~\bibnamefont{Manolescu}},
  \bibnamefont{and}
  \bibinfo{author}{\bibfnamefont{V.}~\bibnamefont{Gudmundsson}},
  \bibinfo{journal}{New J. Phys.} \textbf{\bibinfo{volume}{11}},
  \bibinfo{pages}{073019} (\bibinfo{year}{2009}).

\bibitem[{int()}]{FSC_orderOfMagnitude} \bibinfo{note}{The FSC
    persists beyond the limit of our numerical calculations.}

\bibitem[{\citenamefont{Moldoveanu et~al.}(2010)\citenamefont{Moldoveanu,
  Manolescu, and Gudmundsson}}]{Moldoveanu:2010}
\bibinfo{author}{\bibfnamefont{V.}~\bibnamefont{Moldoveanu}},
  \bibinfo{author}{\bibfnamefont{A.}~\bibnamefont{Manolescu}},
  \bibnamefont{and}
  \bibinfo{author}{\bibfnamefont{V.}~\bibnamefont{Gudmundsson}},
  \bibinfo{journal}{Phys. \ Rev. \ B} \textbf{\bibinfo{volume}{82}},
  \bibinfo{pages}{085311} (\bibinfo{year}{2010}).

\bibitem[{zer()}]{zero_coulomb} \bibinfo{note}{We do not include
    Coulomb interactions among the confined particles. One reason for
    this is that a fully non-Markovian treatment of the dynamics of a
    model with explicit Coulomb interactions introduces an
    extraordinary level of complexity in the calculations, and we do
    not believe it qualitatively affects the Fock-space coherence we
    focus on in this work. For recent Markovian treatments of a model
    with explicit Coulomb interactions see Refs~[28,29]. A second reason
    to neglect this particle-particle interaction is that that our
    results remain valid in configurations where the dot, in
    equilibrium, is occupied by zero or one fermions.}

\bibitem[{tun()}]{tunnel_index} \bibinfo{note}{The tunnelling
    coefficient may be written with an additional index labelling the
    reservoir state. However, for the parameter regime in which we
    focus, the variation of the tunnelling coefficient with dot
    orbitals is much greater than that with the lead states.}

\bibitem[{\citenamefont{Vaz and Kyriakidis}(2010)}]{PhysRevB.81.085315}
\bibinfo{author}{\bibfnamefont{E.}~\bibnamefont{Vaz}} \bibnamefont{and}
  \bibinfo{author}{\bibfnamefont{J.}~\bibnamefont{Kyriakidis}},
  \bibinfo{journal}{Phys. Rev. B} \textbf{\bibinfo{volume}{81}},
  \bibinfo{pages}{085315} (\bibinfo{year}{2010}).

\bibitem[{unp()}]{unpub}
\bibinfo{note}{Eduardo Vaz and Jordan Kyriakidis, unpublished}.

\bibitem[{\citenamefont{Kyriakidis et~al.}(2002)\citenamefont{Kyriakidis,
  Pioro-Ladriere, Ciorga, Sachrajda, and
  Hawrylak}}]{Kyriakidis2002Voltage-tunable}
\bibinfo{author}{\bibfnamefont{J.}~\bibnamefont{Kyriakidis}},
  \bibinfo{author}{\bibfnamefont{M.}~\bibnamefont{Pioro-Ladriere}},
  \bibinfo{author}{\bibfnamefont{M.}~\bibnamefont{Ciorga}},
  \bibinfo{author}{\bibfnamefont{A.~S.} \bibnamefont{Sachrajda}},
  \bibnamefont{and} \bibinfo{author}{\bibfnamefont{P.}~\bibnamefont{Hawrylak}},
  \bibinfo{journal}{Phys.\ Rev.\ B} \textbf{\bibinfo{volume}{66}},
  \bibinfo{pages}{035320} (\bibinfo{year}{2002}).

\bibitem[{rho()}]{rho02} 
\bibinfo{note}{One additional
    element, $\rho_{02}$, describing states differing by two confined
    particles is additionally not depicted.  Such elements form in
    still other isolated blocks.  The overall block diagonal structure
    is maintained for \emph{arbitrary} numbers of transport channels
    differing by arbitrary number of confined particles.}

\bibitem[{int()}]{integration}
\bibinfo{note}{We numerically evaluate the Bromwich integral using an adaptive
  Fourier integration routine and tables of Chebyshev moments. See, for
  example, the \textsc{qawf} algorithms in the GNU scientific library
  (http://www.gnu.org/software/gsl).}

\bibitem[{Teq()}]{Tequal} 
\bibinfo{note}{We impose the
    equality
    $T_{\beta}^{\text{s}}/T_{\alpha}^{\text{s}}=T_{\beta}^{\text{d}}/T_{\alpha}^{\text{d}}$
    solely for reasons of simplicity.}

\bibitem[{\citenamefont{Ciorga et~al.}(2002)\citenamefont{Ciorga, Wensauer,
  Pioro-Ladriere, Korkusinski, Kyriakidis, Sachrajda, and
  Hawrylak}}]{ciorg02:collap.spin.singl}
\bibinfo{author}{\bibfnamefont{M.}~\bibnamefont{Ciorga}},
  \bibinfo{author}{\bibfnamefont{A.}~\bibnamefont{Wensauer}},
  \bibinfo{author}{\bibfnamefont{M.}~\bibnamefont{Pioro-Ladriere}},
  \bibinfo{author}{\bibfnamefont{M.}~\bibnamefont{Korkusinski}},
  \bibinfo{author}{\bibfnamefont{J.}~\bibnamefont{Kyriakidis}},
  \bibinfo{author}{\bibfnamefont{A.~S.} \bibnamefont{Sachrajda}},
  \bibnamefont{and} \bibinfo{author}{\bibfnamefont{P.}~\bibnamefont{Hawrylak}},
  \bibinfo{journal}{Phys.\ Rev.\ Lett.} \textbf{\bibinfo{volume}{88}},
  \bibinfo{pages}{256804} (\bibinfo{year}{2002}).


\bibitem[{\citenamefont{Auston et~al.}(1988)\citenamefont{Auston, Nuss}}]{Auston_Nuss_1988}
\bibinfo{author}{\bibfnamefont{D.}~\bibnamefont{Auston}}, \bibnamefont{and}
  \bibinfo{author}{\bibfnamefont{M.}~\bibnamefont{Nuss}}, 
 \bibinfo{journal}{IEEE J. Quantum Electron.} \textbf{\bibinfo{volume}{24}}
  \bibinfo{pages}{184}  (\bibinfo{year}{2010}).

\bibitem{RevModPhys.83.543}
Ulbricht R, Hendry E, Shan J, Heinz T~F and Bonn M, Rev. Mod. Phys.
  {\bf 83} 543--586 (2011).

\bibitem[{\citenamefont{Cavalleri et~al.}(2011)\citenamefont{Dienst,
      Hoffmann, Fausti, Petersen, Pyon, Takayama, Takagi, and
      Cavalleri}}]{Cavalleri_Nature_2011}
  \bibinfo{author}{\bibfnamefont{A.}~\bibnamefont{Dienst}},
  \bibinfo{author}{\bibfnamefont{M. C.}~\bibnamefont{Hoffmann}},
  \bibinfo{author}{\bibfnamefont{D.}~\bibnamefont{Fausti}},
  \bibinfo{author}{\bibfnamefont{J. C.}~\bibnamefont{Petersen}},
  \bibinfo{author}{\bibfnamefont{S.}~\bibnamefont{Pyon}},
  \bibinfo{author}{\bibfnamefont{T.}~\bibnamefont{Takayama}},
  \bibinfo{author}{\bibfnamefont{H.}~\bibnamefont{Takagi}},
  \bibnamefont{and}
  \bibinfo{author}{\bibfnamefont{A.}~\bibnamefont{Cavalleri}},
  \bibinfo{journal}{Nature Photon.} \textbf{\bibinfo{volume}{5}},
  \bibinfo{number}{8} \bibinfo{pages}{485--488}
  (\bibinfo{year}{2010}).

\bibitem[{\citenamefont{Miyamoto_2005}(2005)}]{Miyamoto_2005}
  \bibinfo{author}{\bibfnamefont{M.}~\bibnamefont{Miyamoto}},
  \bibinfo{journal}{Phys. Rev. A} \textbf{\bibinfo{volume}{72}},
  \bibinfo{pages}{063405} (\bibinfo{year}{2005}).

\bibitem[{\citenamefont{Bulgakov_2007}(2007)}]{Bulgakov_2007}
  \bibinfo{author}{\bibfnamefont{E.~N.}~\bibnamefont{Bulgakov}},
  \bibinfo{author}{\bibfnamefont{I.}~\bibnamefont{Rotter}},
  \bibnamefont{and}
  \bibinfo{author}{\bibfnamefont{A.~F.}~\bibnamefont{Sadreev}},
  \bibinfo{journal}{Phys. Rev. A } \textbf{\bibinfo{volume}{75}},
  \bibinfo{pages}{067401} (\bibinfo{year}{2007}).

\bibitem[{\citenamefont{Tong_2010}(2010)}]{Tong_2010}
  \bibinfo{author}{\bibfnamefont{Q.~J.}~\bibnamefont{Tong}},
  \bibinfo{author}{\bibfnamefont{J.~H.}~\bibnamefont{An}},
  \bibinfo{author}{\bibfnamefont{H.~G.}~\bibnamefont{Luo}},
  \bibnamefont{and}
  \bibinfo{author}{\bibfnamefont{C.~H.}~\bibnamefont{Oh}},
  \bibinfo{journal}{Phys. Rev. A} \textbf{\bibinfo{volume}{81}},
  \bibinfo{pages}{052330} (\bibinfo{year}{2010}).

\bibitem[{\citenamefont{Sadreev_2006}(2006)}]{Sadreev_2006}
  \bibinfo{author}{\bibfnamefont{A.~F.}~\bibnamefont{Sadreev}},
  \bibinfo{author}{\bibfnamefont{E.~N.}~\bibnamefont{Bulgakov}},
  \bibnamefont{and}
  \bibinfo{author}{\bibfnamefont{I.}~\bibnamefont{Rotter}},
  \bibinfo{journal}{Phys. Rev. B } \textbf{\bibinfo{volume}{73}},
  \bibinfo{pages}{235342} (\bibinfo{year}{2006}).

\bibitem[{\citenamefont{Magunov_1999}(1999)}]{Magunov_1999}
  \bibinfo{author}{\bibfnamefont{A.~I.}~\bibnamefont{Magunov}},
  \bibinfo{author}{\bibfnamefont{I.}~\bibnamefont{Rotter}},
  \bibnamefont{and}
  \bibinfo{author}{\bibfnamefont{S.~I.}~\bibnamefont{Strakhova}},
  \bibinfo{journal}{J.~Phys. B} \textbf{\bibinfo{volume}{32}},
  \bibinfo{pages}{1669} (\bibinfo{year}{1999}).

\bibitem[{\citenamefont{Rotter_2005}(2005)}]{Rotter_2005}
  \bibinfo{author}{\bibfnamefont{I.}~\bibnamefont{Rotter}},
  \bibnamefont{and}
  \bibinfo{author}{\bibfnamefont{A.~F.}~\bibnamefont{Sadreev}},
  \bibinfo{journal}{Phys. Rev. E} \textbf{\bibinfo{volume}{71}},
  \bibinfo{pages}{046204} (\bibinfo{year}{2005}).

\bibitem[{\citenamefont{vaz_kyriakidis_2008}(2008)}]{vaz_kyriakidis_2008}
  \bibinfo{author}{\bibfnamefont{E.}~\bibnamefont{Vaz}},
  \bibnamefont{and}
  \bibinfo{author}{\bibfnamefont{J.}~\bibnamefont{Kyriakidis}},
  \bibinfo{journal}{J.~Chem.~Phys.} \textbf{\bibinfo{volume}{129}},
  \bibinfo{pages}{024703} (\bibinfo{year}{2008}).

\bibitem[{\citenamefont{Esposito et~al.}(2009)\citenamefont{Esposito,
      Massimiliano and Galperin, Michael}}]{Esposito:2009}
  \bibinfo{author}{\bibfnamefont{M.}~\bibnamefont{Esposito}},
  \bibnamefont{and}
  \bibinfo{author}{\bibfnamefont{M.}~\bibnamefont{Galperin}},
  \bibinfo{journal}{Phys. \ Rev. \ B} \textbf{\bibinfo{volume}{79}},
  \bibinfo{pages}{205303} (\bibinfo{year}{2009}).

\bibitem[{\citenamefont{Bulnes et~al.}(2011)\citenamefont{Bulnes
      Cuetara, Gregory and Esposito, Massimiliano and Gaspard,
      Pierre}}]{Bulnes:2011}
  \bibinfo{author}{\bibfnamefont{G.}~\bibnamefont{Bulnes Cuetara}},
  \bibinfo{author}{\bibfnamefont{M.}~\bibnamefont{Esposito}},
  \bibnamefont{and}
  \bibinfo{author}{\bibfnamefont{P.}~\bibnamefont{Gaspard}},
  \bibinfo{journal}{Phys. \ Rev. \ B} \textbf{\bibinfo{volume}{84}},
  \bibinfo{pages}{165114} (\bibinfo{year}{2011}).

\end{thebibliography}
\end{document}